\documentclass[runningheads]{llncs}
\usepackage{amsmath}
\usepackage{amsfonts}
\usepackage{wrapfig}
\usepackage{hyperref}
\usepackage{color}

\usepackage{bbm}
\usepackage{graphicx}

\newcommand{\xdownarrow}[1]{%
	{\left\downarrow\vbox to #1{}\right.\kern-\nulldelimiterspace}
}
\usepackage{fancyhdr}

\fancypagestyle{firstpage}
{
	\fancyhead[L]{}    
	\fancyhead[C]{\textit{\small }}
}
\begin{document}

	\title{Effects of noise on leaky integrate-and-fire neuron models for neuromorphic computing applications}
	\titlerunning{Effects of noise on a leaky integrate-and-fire neuron models for neuromorphic computing applications}
	
	\author{Thi Kim Thoa Thieu\inst{1} \and
		Roderick Melnik\inst{1,2}}
	%Author\inst{1}\orcidID{0000-1111-2222-3333} \and
	%Second Author\inst{2,3}\orcidID{1111-2222-3333-4444} \and
	%Third Author\inst{3}\orcidID{2222--3333-4444-5555}}
	%
	\authorrunning{Thoa Thieu and Roderick Melnik}
	\institute{MS2Discovery Interdisciplinary Research Institute, Wilfrid Laurier University, \\75 University Ave W, Waterloo, Ontario, Canada N2L 3C5 
		\\
		\and
		BCAM - Basque Center for Applied Mathematics, Bilbao, Spain\\ \email{\{tthieu, rmelnik\}@wlu.ca}
	}
	%\institute{MS2Discovery Interdisciplinary Research Institute, Wilfrid Laurier University, 75 University Ave W, Waterloo, Ontario, Canada N2L 3C5 
	%	\email{lncs@springer.com}\\
	%	\url{http://www.springer.com/gp/computer-science/lncs} \and
	%	ABC Institute, Rupert-Karls-University Heidelberg, Heidelberg, Germany\\
	%	\email{\{abc,lncs\}@uni-heidelberg.de}}
	%
	\maketitle              % typeset the header of the contribution
	%
	%\footnotesize \textit{\small The paper is under review in the Proceedings of ICCS-2021, Springer, 2021}
	\thispagestyle{firstpage}
	\begin{abstract}
		
		Artificial neural networks (ANNs) have been extensively used for the description of problems arising from biological systems and for constructing neuromorphic computing models. The third generation of ANNs, namely, spiking neural networks (SNNs), inspired by biological neurons enable a more realistic mimicry of the human brain. A large class of the problems from these domains is characterized by the necessity to deal with the combination of neurons, spikes and synapses via integrate-and-fire neuron models. Motivated by important applications of the integrate-and-fire of neurons in neuromorphic computing for bio-medical studies, the main focus of the present work is on the analysis of the effects of additive and multiplicative types of random input currents together with a random refractory period on a leaky integrate-and-fire (LIF) synaptic conductance neuron model. Our analysis is carried out via Langevin stochastic dynamics in a numerical setting describing a cell membrane potential. We provide the details of the model, as well as representative numerical examples, and discuss the effects of noise on the time evolution of the membrane potential as well as the spiking activities of neurons in the LIF synaptic conductance model scrutinized here. Furthermore, our numerical results demonstrate that the presence of a random refractory period in the LIF synaptic conductance system may substantially influence an increased irregularity of spike trains of the output neuron. 
	\end{abstract}
	
	%Motivated by important applications of the integrate-and-fire of neurons in the modelling of neuronal dynamics and spiking activities with artificial neuron and neuromorphic computing in bio-medical study, the main focus of the present work is on the analysis of the effects of random inputs on leaky integrate and fire (LIF) neuron models.
	\section{Introduction}
	%The relationship between reflecting SDEs
	%and some models of queueing theory, which is a subject of very intensive investigations.
	%
	%In the case of coarse-graining and modelling error, the representation is used for approximating Kinetic Monte Carlo jump dynamics by SDE dynamics.
	%
	%The aim of this section is to show a possible relationship between reflecting SDEs and some models of queueing theory, which is a subject of very intensive investigations. We consider a limit theorem for one simple model of queueing theory that leads us to a
	%reflection SDE and discuss possible generalizations.
	
	In recent years, the modelling with artificial neural networks (ANNs) offers many challenging questions to some of the most advanced areas of science and technology \cite{Chowdhury2021}.  The progress in ANNs has led to improvements in various cognitive tasks and tools for vision, language, behavior and so on.  Moreover, some ANN models together with the numerical algorithms bring the outcome achievements at the human-level performance. In general, biological neurons in the human brain transmit information by generating spikes. To improve the biological plausibility of the existing ANNs, spiking neural networks (SNNs) are known as the third generation of ANNs. SNNs play an important role in the modelling of important systems in neuroscience since SNNs more realistically mimic the activity of biological neurons by the combination of neurons and synapses \cite{Chen2021}. In particular, neurons in the SNNs transmit information only when a membrane potential, i.e. an intrinsic quality of the neuron related to its membrane electrical charge, reaches a specific threshold value. The neuron fires, and generates a signal that travels to other neurons when the membrane reaches its threshold. Hence, a neuron that fires in a membrane potential model at the moment of threshold crossing is called a spiking neuron. Many models have been proposed to describe the spiking activities of neurons in different scenarios. One of the simplest models, providing a foundation for many neuromorphic applications, is a leaky integrate-and-fire (LIF) neuron model \cite{Jaras2021,Nandakumar2020,Pottelbergh2021}. The LIF model mimics the dynamics of the cell membrane in the biological system \cite{Cavallari2014,Latimer2019}  and provides a suitable compromise between complexity and analytical tractability when implemented for large neural networks. Recent works have demonstrated the importance of the LIF model that has become one of the most popular neuron models in neuromorphic computing \cite{Brigner2020,Fardet2020,Dutta2017,Gerum2021,Guo2021,Teeter2018}.  However, ANNs are intensively computed and often deal with many challenges from severe accuracy degradation if the testing data is corrupted with noise \cite{Chowdhury2021,Hendrycks2019}, which may not be seen during training. Uncertainties coming from different sources \cite{Faisal2008}, e.g. inputs, devices, chemical reactions, etc would need to be accounted for. Furthermore, the presence of fluctuations can effect on the transmission of a signal in nonlinear systems \cite{Burkitt2006,Burkitt2006-2}. Recent results provided in \cite{Bauermann2019} have shown that multiplicative noise is beneficial for the transmission of sensory signals in
	simple neuron models. To get closer to the real scenarios in biological systems as well as in their computational studies, we are interested in evaluating the contribution of uncertainty factors arising in LIF systems. In particular, we investigate the effects of the additive and multiplicative noise input currents together with the random refractory period on the dynamics of a LIF synaptic conductance system. A better understanding of random input factors in LIF synaptic conductance models would allow for a more efficient usage of smart SNNs and/or ANNs systems in such fields as biomedicine and other applications \cite{Chowdhury2021,Woo2021}. 
	
%	\\[3pt]
	Motivated by LIF models and their applications in SNNs and ANNs subjected to natural random factors in the description of biological systems, we develop a LIF synaptic conductance model of neuronal dynamics to study the effects of additive and multiplicative types of random external current inputs together with a random refractory period on the spiking activities of neurons in a cell membrane potential setting. Our analysis focuses on considering a Langevin stochastic equation in a numerical setting for a cell membrane potential with random inputs. We provide numerical examples and discuss the effects of random inputs on the time evolution of the membrane potential as well as the spiking activities of neurons in our model.  Furthermore, the model of LIF synaptic conductances is examined on the data from dynamic clamping (see, e.g., \cite{Teka2014,Latimer2019,Woo2021}) in the Poissonian input spike train setting. 

	\section{ Random factors and a LIF synaptic conductance neuron model} \label{Sec:classicalLIF}
	\subsection{SNN algorithm and a LIF synaptic conductance neuron model}
	
\begin{figure}[h!]
		\centering
		\includegraphics[width=0.75\textwidth]{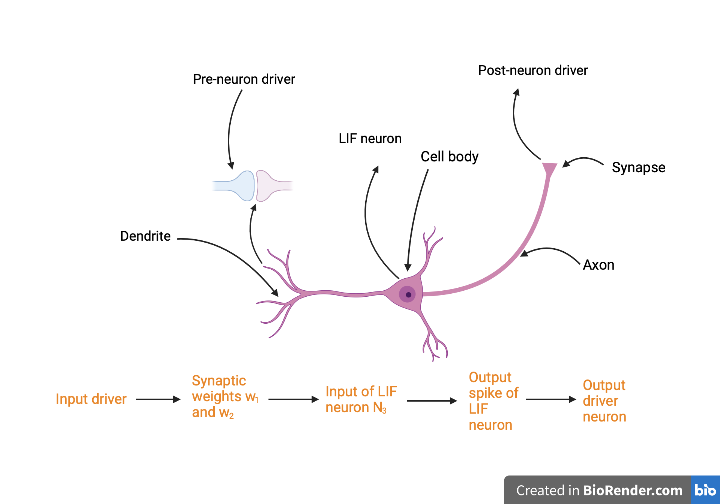}
		\caption{[Color online] Sketch of SNN algorithm.}
		\label{fig:00-1}
	\end{figure}

	Let us recall the SNN algorithm, presented schematically in Fig. \ref{fig:00-1} (see, e.g., \cite{Dutta2017}). At the first step, pre-synaptic neuronal drivers provide the input voltage spikes. Then, we convert the input driver for spikes to a gently varying current signal proportional to the synaptic weights $w_1$ and $w_2$. Next, the synaptic current response is summed into the input of LIF neuron $N_3$. Then, the LIF neuron integrates the input current across a capacitor, which raises its potential. After that, $N_3$ resets immediately (i.e. loses stored charge) once the potential reaches/exceeds a threshold. Finally, every time $N_3$ reaches the threshold, a driver neuron D3 produces a spike.

	In general, the biological neuronal network is related to the SNN algorithm. Moreover, 
	the main role of SNNs is to understand and mimic human brain functionalities since SNNs enable to approximate efficient learning and recognition tasks in neuro-biology. Hence, to have a better implementation of SNNs in hardware, it would be necessary to describe an efficient analog of the biological neuron. Therefore, in what follows, we are interested in the SNN algorithm starting from the third step, where the synaptic current response is summed into the input of LIF neuron, to the last step of the SNN algorithm. In particular, at the third step of SNN algorithm, it is assumed that the summation of synaptic current responses can be a constant, in a deterministic form or can be even represented by a random type of current. To get closer to the real scenarios of neuronal models, we should also account for the existence of random fluctuations in the systems. Specifically, the random inputs arise primarily through sensory fluctuations, brainstem discharges and thermal energy (random fluctuations at a microscopic level, such as Brownian motions of ions). The stochasticity can arise even from the devices which are used for medical treatments, e.g. devices for injection currents into the neuronal systems. For simplicity, we consider a LIF synaptic conductance model with additive and multiplicative noise input currents in presence of a random refractory period. 
	In biological systems such as brain networks, instead of physically joined neurons, a spike in the presynaptic cell causes a chemical, or a neurotransmitter, to be released into a small space between the neurons called the synaptic cleft \cite{Gerstner2014}. Therefore, in what follows, we will focus on investigating chemical synaptic transmission and study how excitation and inhibition affect the patterns in the neurons' spiking output. In this section, we consider a model of synaptic conductance dynamics. In particular, neurons receive a myriad of excitatory and inhibitory synaptic inputs at dendrites. To better understand the mechanisms of synaptic conductance dynamics, we investigate the dynamics of the random excitatiory (E) and inhibitory inputs  to a neuron \cite{Li2019}.

	In general, synaptic inputs are the combination of excitatory neurotransmitters. Such neurotransmitters depolarize the cell and drive it towards the spike threshold, while inhibitory neurotransmitters hyperpolarize it and drive it away from the spike threshold. These chemical factors cause specific ion channels on the postsynaptic neuron to open. Then, the results make a change in the neuron's conductance. Therefore, the current will flow in or out of the cell \cite{Gerstner2014}.
	
	For simplicity, we define transmitter-activated ion channels as an explicitly time-dependent conductivity $(g_{\text{syn}}(t))$. Such conductance transients can be generated by the following equation (see, e.g., \cite{Dayan2005,Gerstner2014}): 
	
	\begin{align}\label{conductivity}
		\frac{d g_{\text{syn}}(t)}{dt} = -\bar{g}_{\text{syn}}\sum_{k}\delta(t-t_k) - \frac{g_{\text{syn}}(t)}{\tau_{\text{syn}}},
	\end{align}
	where $\bar{g}_{\text{syn}}$ (synaptic weight) is the maximum conductance elicited by each incoming spike, while $\tau_{\text{syn}}$ is the synaptic time constant and $\delta(\cdot)$ is the Dirac delta function. Note that the summation runs over all spikes received by the neuron at time $t_k$. Using Ohm's law, we have the following formula for converting conductance changes to the current:
	
	\begin{align}
		I_{\text{syn}}(t) = g_{\text{syn}}(t)(V(t) - E_{\text{syn}}), 
	\end{align}
	where $E_{\text{syn}}$ represents the direction of current flow and the excitatory or inhibitory nature of the synapse.

	%Thus, incoming spikes are filtered by an exponential-shaped kernel, effectively low-pass filtering the input. In other words, synaptic input is not white noise, but it is, in fact, colored noise, where the color (spectrum) of the noise is determined by the synaptic time constants of both excitatory and inhibitory synapses.
	
	In general, the total synaptic input current $I_{\text{syn}}$ is the sum of both excitatory and inhibitory inputs. We assume that the total excitatory and inhibitory conductances received at time $t$ are $g_E(t)$ and $g_I(t)$, and their corresponding reversal potentials are $E_E$ and $E_I$, respectively.  We define the total synaptic current by the following equation: 
	
	\begin{align}
		I_{\text{syn}}(V(t),t) = -g_E(t) (V-E_E) - g_I(t)(V-E_I). 
	\end{align}
	
	Therefore, the corresponding membrane potential dynamics of the LIF neuron under synaptic current (see, e.g., \cite{Li2019}) can be described as follows:

	\begin{align}\label{main_eq2}
		\tau_m \frac{d}{dt}V(t) &= -( V(t) - E_L)  - \frac{g_E(t)}{g_L}(V(t) - E_E) -  \frac{g_I(t)}{g_L}(V(t) - E_I) + \frac{I_{\text{inj}}}{g_L}, 
	\end{align}
	where $V$ is the membrane potential, $I_{\text{inj}}$ is the external input current, while $\tau_m$ is the membrane time constant. We consider the membrane potential model where a spike takes place whenever $V(t)$ crosses $V_{\text{th}}$. Here, $V_{\text{th}}$ denotes the membrane potential threshold to fire an action potential. In that case, a spike is recorded and $V(t)$ resets to $V_{\text{reset}}$ value. This is summarized in the reset condition $V(t) = V_{\text{reset}}$ if $V(t) \geq V_{\text{th}}$. We define the following LIF model with and a reset condition: 
	
	\begin{align}
		\tau_m \frac{d}{dt}V(t) &= -( V(t) - E_L)  - \frac{g_E(t)}{g_L}(V(t) - E_E) -  \frac{g_I(t)}{g_L}(V(t) - E_I) \nonumber\\&+ \frac{I_{\text{inj}}}{g_L} \quad \text{ if } V(t) \leq V_{\text{th}}, \\
		V(t) &= V_{\text{reset}} \quad \text{ otherwise},
	\end{align}
	In this model, we consider a random synaptic input by introducing the following random input current (additive noise)
		$I_{\text{inj}} = I_0 + \sigma_1 \eta(t)$, where $\eta$ is the zero-mean Gaussian white noise with unit variance. For the multiplicative noise case, the applied current is set to $I_{\text{inj}} = V(t)(I_0 + \sigma_2 \eta(t))$. Here, $\sigma_1, \sigma_2$ denote the standard deviations of these random components to the inputs. When considering such random input currents, the equation \eqref{main_eq2} can be considered as the following Langevin stochastic equation (see, e.g., \cite{Roberts2017}):
		
		\begin{align}\label{main_eq02}
\tau_m \frac{d}{dt}V(t) &= -( V(t) - E_L)  - \frac{g_E(t)}{g_L}(V_m(t) - E_E) -  \frac{g_I(t)}{g_L}(V_m(t) - E_I) \nonumber\\&+ \begin{cases}\frac{1}{g_L}(I_0 + \sigma_1 \eta(t))\\
				\frac{1}{g_L}V(t)(I_0 + \sigma_2 \eta(t))
			\end{cases} \quad \text{ if } V(t) \leq V_{\text{th}}. 
		\end{align}
%	Similar to the first model analyzed in Section \ref{Sec:classicalLIF}, we consider here random synaptic inputs by introducing the following random input current (additive noise)
%	$I_{\text{inj}} = I_0 + \sigma_1 \eta(t)$, where $\eta$ is the zero-mean Gaussian white noise with unit variance. In this representation, $\sigma_1$ denotes the standard deviation of this random component to the input. For the multiplicative noise case, the applied current is set to $I_{\text{inj}} = I_0 + \sigma_2 V(t)\eta(t)$, where $\sigma_2$ represents the standard deviation of this random components to the input. In what follows, we also consider the random refractory period as in the previous section. Similarly, in the presence of such random input currents, the equation \eqref{main_eq2} can be seen as a Langevin stochastic equation. 
	
	In our model, we use the simplest input spikes with Poisson process 
	which provide a suitable approximation to stochastic neuronal firings \cite{Kuhn2004}. This input spikes will be added in the quantity $\sum_{k}\delta(t-t_k)$ in the equation \eqref{conductivity}. In particular, the input spikes are given when every input spike arrives independently of other spikes. For designing a spike generator of spike train, let us call the probability of firing a spike within a short interval  (see, e.g. \cite{Dayan2005}) $P(1 \text{ spike during } \Delta t) = r_{j}\Delta t$, where $j=e,i$ with $r_e, r_i$ representing the instantaneous excitatory and inhibitory firing rates, respectively. This expression is designed to generate a Poisson spike train by first subdividing time into a group of short intervals through small time steps $\Delta t$. At each time step, we define a random variable $x_{\text{rand}}$ with uniform distribution
	over the range between 0 and 1. Then, we compare this with the probability of firing a spike, which is described as follows:
	
	\begin{align}
		\begin{cases}
			r_j\Delta t > x_{\text{rand}}, \text{ generates a spike},\\
			r_j \Delta t \leq x_{\text{rand}}, \text{ no spike
				is generated}.
		\end{cases}
	\end{align}
	
%	We know that there are three main events taking place during an action potential, namely, depolarization, repolarization and hyperpolarization. The action potential frequency shows how often action potentials are sent. There exists a maximum frequency at which a single neuron can send action potentials, and this is determined by its refractory periods. Hence, the absolute refractory period is a time interval on the order of a few milliseconds during which a synaptic input will not lead to a second spike, see e.g. \cite{Mahdi2013}.
	In this work, we also investigate the effects of random refractory periods \cite{Mahdi2013}. We define the random refractory periods $t_{\text{ref}}$ as $t_{\text{ref}} = \mu_{\text{ref}} + \sigma_{\text{ref}}\tilde{\eta}(t)$, where $ \tilde{\eta}(t) \sim \mathcal{N}(0,1)$. 
	
%	We have demonstrated the importance of effects of random inputs on the classical LIF system. However, 

	%The Poisson process provides
	%an extremely useful approximation of stochastic neuronal firing. To make Poisson process
	%the presentation easier to follow, we separate two cases, the homogeneous
	%Poisson process, for which the firing rate is constant over time, and the
	%inhomogeneous Poisson process, which involves a time-dependent firing
	%rate.
	\subsection{Firing rate and spike time irregularity }
	
	In general, the irregularity of spike trains can provide information about  stimulating activities in a neuron. A LIF synaptic conductance neuron with multiple inputs and coefficient of variation (CV) of the inter-spike-interval (ISI) can bring an output decoded neuron. In this work, we show that the increase $\sigma_{\text{ref}}$ can lead to an increase in the irregularity of the spike trains (see also \cite{Gallinaro2021}).

	We define the spike regularity via coefficient of variation of the inter-spike-interval (see, e.g., \cite{Christodoulou2001,Gallinaro2021}) as follows:
	$$CV_{\text{ISI}} = \frac{\sigma_\text{ISI}}{\mu_\text{ISI}},$$
	where $\sigma_\text{ISI}$ is the standard deviation and $\mu_\text{ISI}$ is the mean of the ISI of an individual neuron.

	In the next section, we consider the output firing rate as a function of Gaussian white noise mean or direct current value, known as the input-output transfer function of the neuron.

	\section{Numerical results for the LIF synaptic conductance model}\label{num-model2}
	
	In this subsection, we take a single neuron at the dendrite and study how the neuron behaves when it is bombarded with both excitatory and inhibitory spike trains (see, e.g., \cite{Latimer2019,Li2019}). 
	
	The simulations this section have been carried out by using by a discrete-time integration based on the Euler method inplemented in Python.
%	a modification of the numerical method provided in the open source framework at \url{https://github.com/} (see W2D3 Biological Neuron Models in the Neuromatch Academy directory).
	
\begin{figure}[h!]
		\centering
		\includegraphics[width=1.0\textwidth]{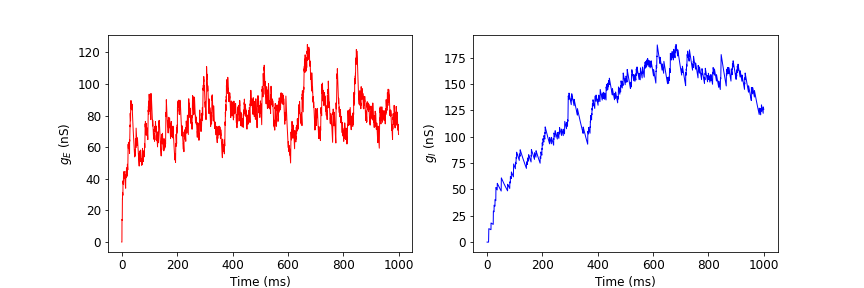} 
		\caption{[Color online]  Left: Excitatory conductances profile. Right: Inhibitory conductances profile.}
		\label{fig:1-0}
	\end{figure}

	In the simulations, we choose the parameter set as follows: $E_E = 70$ (mV), $E_L = -60$ (mV), $E_I = -10$ (mV), $V_{\text{th}} = -55$ (mV), $V_{\text{reset}} = -70$ (mV), $\Delta t = 0.1$, $\tau_m = 10$ (ms), $\tau_{E} =20$ (ms), $\tau_{I} = 100$ (ms), $\bar{g}_E = 4.8$ (nS), $\bar{g}_I = 6.4$(nS), $r_e = 10$, $r_i = 10$, $n_E = 80$ spikes, $n_I = 20$ spikes. Here, $n_E$ and $n_I$ represent the number of excitatory and inhibitory presynaptic spike trains, respectively. These parameters have also been used in \cite{Latimer2019,Li2019} for dynamic
	clamp experiments and we take them for our model validation. In this subsection, we use the excitatory and inhibitory conductances provided in Fig. \ref{fig:1-0} for all of our simulations. 

	The main numerical results of our analysis here are shown in Figs.\ref{fig:0}-\ref{fig:1-4}, where we have plotted the time evolution of the membrane potential calculated based on model \eqref{main_eq2}, the distribution of the ISI and the corresponding spike irregularity profile. We investigate the effects of additive and multiplicative types of random input currents inpresence of a random refractory period on a LIF neuron under synaptic conductance dynamics. Under a Poissonian spike input, the random external currents and random refractory period influence the spiking activity of a neuron in the cell membrane potential. 

	\subsection{Additive noise}
	
	 \begin{figure}[h!]
		\centering
			\includegraphics[width=0.8\textwidth]{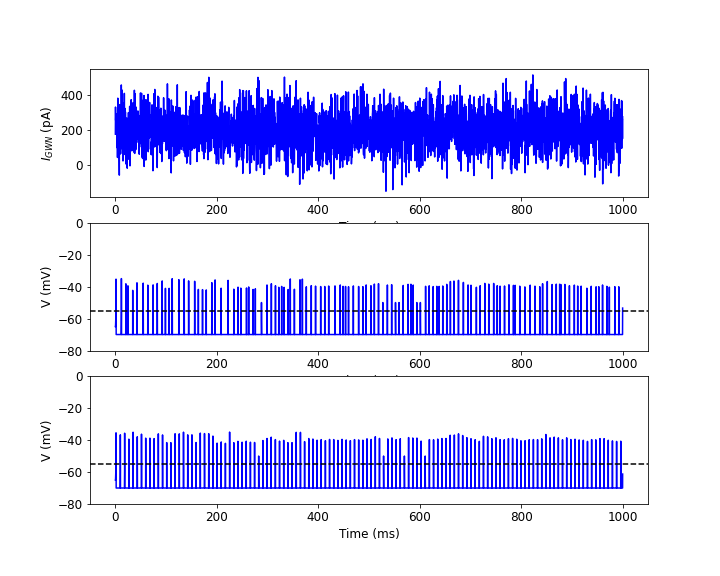} 
		\caption{[Color online] Top row: Gaussian white noise current profile. Middle row: Time evolution of membrane potential $V(t)$ with additive noise current and random refractory period $t_{\text{ref}} = 8 + 2\tilde{\eta}(t)$ (ms). Bottom row: Time evolution of membrane potential $V(t)$ with direct input current and direct refractory period $t_{\text{ref}} = 8 $ (ms). The dash line represents the spike threshold $V_{\text{th}} = -55$ (mV).
	%		TThe circuit diagram (left) represents the schematic layout of the integrate-and-firecomponents. The graph (right) depicts voltage vs time for a neuronstimulated by a constant current.
	}\label{fig:0}	
	\end{figure}
In Fig. \ref{fig:0}, we have plotted the Gaussian white noise current profile, the time evolution of the membrane potential $V(t)$ with Gaussian white noise input current ($I_{\text{inj}} = 200 + \eta(t)$ (pA)) and direct input current ($I_{\text{inj}} = I_{\text{dc}} = 200$ (pA)). In this case, we fix the value of $t_{\text{ref}} = 8 + 2\tilde{\eta}(t)$ (ms).  We observe that the time evolution of the membrane potential looks quite similar in the two cases. Note that a burst occurred when a neuron spiked more than once within 25 (ms) (see, e.g., \cite{So2012}). In this case, when considering the presence of random input current and random refractory period in the system, we observe there exist bursts in the case presented in the second row of Fig. \ref{fig:0}.
%There is an increase in the distance between each spike in both cases. 
\begin{figure}[h!]
	\centering
	\begin{tabular}{ll}
	\includegraphics[width=0.5\textwidth]{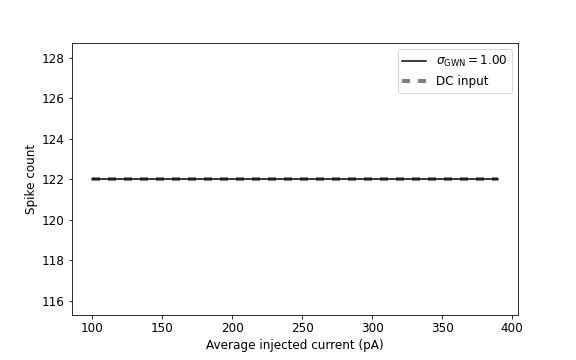} &\includegraphics[width=0.5\textwidth]{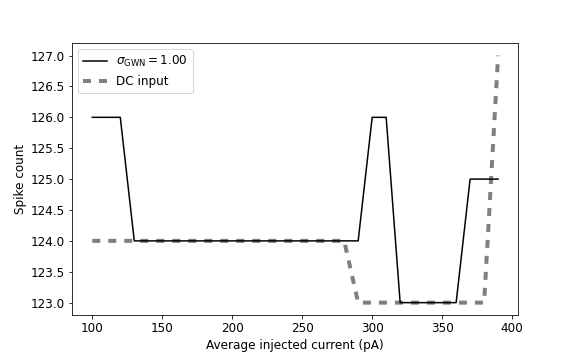} 
\end{tabular}
	\caption{[Color online] The input-output transfer function of the neuron, output firing rate as a function of input mean for the case with additive noise input current ( $\sigma_{\text{Inj}} = 1$).  Left: direct time refractory period $t_{\text{ref}} = 8$ (ms). Right: random refractory period $t_{\text{ref}} = 8 + 2\tilde{\eta}(t)$ (ms).
		%		TThe circuit diagram (left) represents the schematic layout of the integrate-and-firecomponents. The graph (right) depicts voltage vs time for a neuronstimulated by a constant current.
	}\label{fig:0-1}	
\end{figure}

We look also at the input-output transfer function of the neuron, the output firing as a function of average injected current in Fig. \ref{fig:0-1}. In particular, we see that the spike count values are slightly fluctuating when we add the random refractory period into the system (in the right panel of Fig. \ref{fig:0-1}). Moreover, in the averaged injected current intervals $[130;290]$ (pA) and $[320;360]$ (pA), the spike count value is the same in both cases: the Gaussian white noise and direct currents in the right panel of Fig. \ref{fig:0-1}. We have seen this phenomenon for $I_{\text{inj}} = 200$ (pA) also in the Fig. \ref{fig:0}.
	\begin{figure}[h!]
	\centering
	\begin{tabular}{l}
		\includegraphics[width=0.8\textwidth]{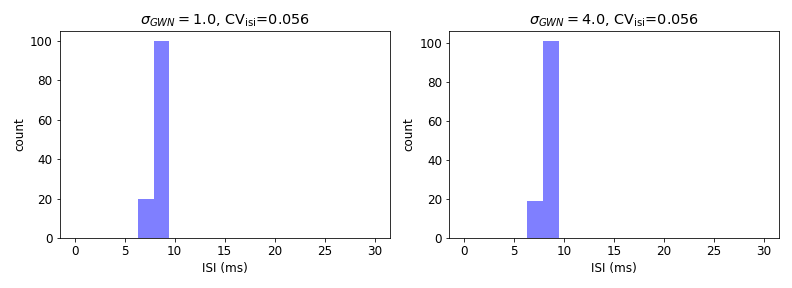} \\
		\includegraphics[width=0.8\textwidth]{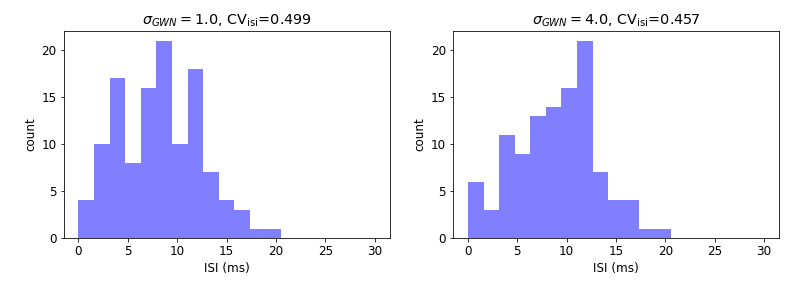}
	\end{tabular}
	\caption{[Color online] ISI histogram distributions for the case with additive noise input current. First row: $t_{\text{ref}} = 8 + 0.5\tilde{\eta}(t)$ (ms). Second row: $t_{\text{ref}} = 8 + 4\tilde{\eta}(t)$ (ms).}
	\label{fig:0-2}
\end{figure}

\begin{figure}[h!]
	\centering
		\begin{tabular}{ll}
	\includegraphics[width=0.4\textwidth]{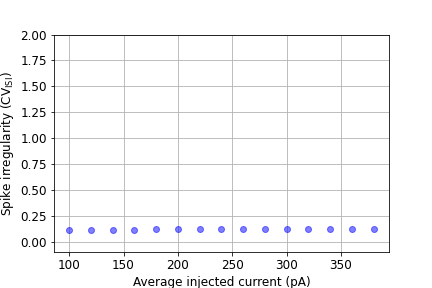} & 	\includegraphics[width=0.4\textwidth]{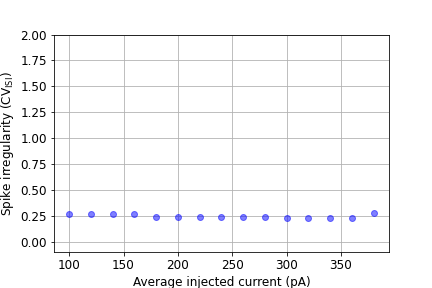}\\
	\includegraphics[width=0.4\textwidth]{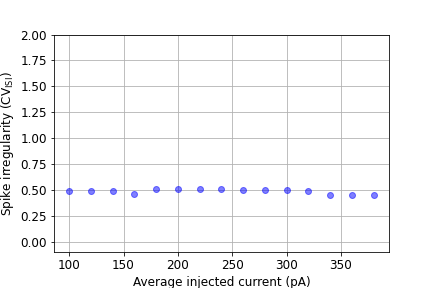} & 	\includegraphics[width=0.4\textwidth]{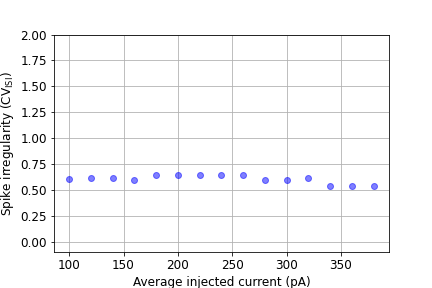}
		\end{tabular}
	\caption{[Color online] Spike irregularity profile in the case with additive noise input current. Top left panel: $t_{\text{ref}} = 8 + \tilde{\eta}(t)$ (ms). Top right panel: $t_{\text{ref}} = 8 + 2\tilde{\eta}(t)$ (ms). Bottom left panel: $t_{\text{ref}} = 8 + 4\tilde{\eta}(t)$ (ms). Bottom right panel: $t_{\text{ref}} = 8 + 6\tilde{\eta}(t)$ (ms).
		%		TThe circuit diagram (left) represents the schematic layout of the integrate-and-firecomponents. The graph (right) depicts voltage vs time for a neuronstimulated by a constant current.
	}\label{fig:0-3}	
\end{figure}

In Fig.\ref{fig:0-2}, we have plotted the spike count distribution as a function of ISI. We observe that the coefficient $CV_{\text{isi}}$ increases when we increase the value of $\sigma_{\text{ref}}$. The spikes are distributed almost entirely in the ISI interval from 6 to 9 (ms) in the first row, while the spikes are distributed mostly in from 0 to 21 (ms) in the second row of  Fig.\ref{fig:0-2}. The ISI distribution presented in the first row of Fig.\ref{fig:0-2} reflects bursting moods. To understand better such phenomenon let us look at the following spike irregularity profile in Fig. \ref{fig:0-3}. 
In Fig.\ref{fig:0-3}, we look at the corresponding spike irregularity profile of the spike count in Figs. \ref{fig:0-2}-\ref{fig:0-3}.  In this plot, we fix the external current $I_{\text{inj}} = 200 + \eta(t)$ (pA) together with considering different values of $\sigma_{\text{ref}}$. We observe that when we increase the value of $\sigma_{\text{ref}}$ the coefficient $CV_{\text{isi}}$ increases. In general, when we increase the mean of the Gaussian white noise, at some point, the
effective input means are above the spike threshold and then the neuron operates in the so-called mean-driven regime. Hence, as the input is sufficiently high, the neuron is charged up to the spike threshold and then it is reset. This essentially gives an almost regular spiking. However, in our case, by considering various values of the random refractory period, we see that $CV_{\text{isi}}$ increases when we increase the values of $\sigma_{\text{ref}}$. This is visible in Fig. \ref{fig:0-3}, $CV_{\text{isi}}$ increases from 0.1 to 0.6 when $\sigma_{\text{ref}}$ increases from 1 to 6. Note that an increased ISI regularity could result in bursting \cite{Maimon2009}. Moreover, the spike trains are substantially more regular with a range $CV_{\text{ISI}} \in (0; 0.5)$, and more irregular when $CV_{\text{ISI}} > 0.5$ \cite{Stiefel2013}. Therefore, in some cases, the presence of random input current with oscilations could lead to the burst discharge. 

%The presence of a random refractory period increases the distance of the time interval between two nearest neighbor spikes as well as decreases the spiking activity in the system.
%This is visible also in the corresponding spike irregularity profile in Fig. \ref{fig:9}, at the average injected current of value 250 (pA), we see a decrease of the spike irregularity coefficient $\text{CV}_{\text{ISI}}$ from 1.7 to 1.1. It is clear that even with a decrease in the spike irregularity the coefficient $\text{CV}_{\text{ISI}}$, in this case, is still larger than in the cases presented in Fig. \ref{fig:7}. This is due to the fact that when we increase the mean of the Gaussian white noise, at some point, the
%effective input means are above the spike threshold and then the neuron operates in the so-called mean-driven regime. Hence, as the input is sufficiently high, the neuron is charged up to the spike threshold and then it is reset. This essentially gives an almost regular spiking.
	\subsection{Multiplicative noise}
	
		 \begin{figure}[h!]
		\centering
		\includegraphics[width=0.7\textwidth]{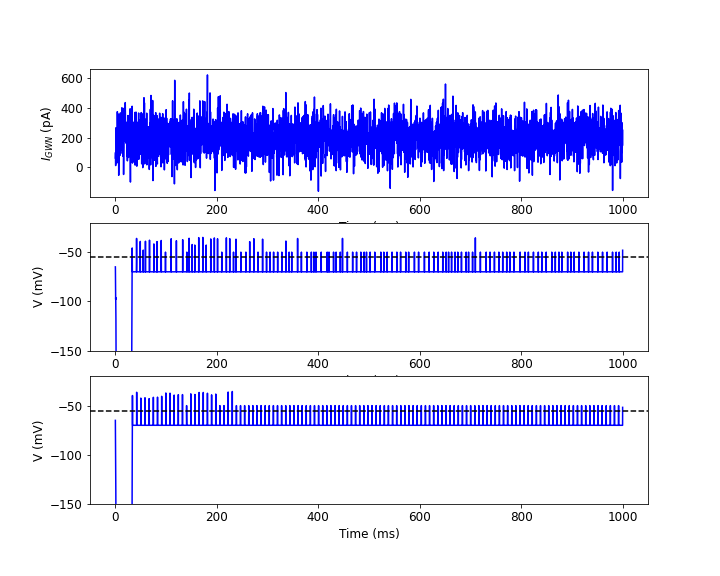} 
		\caption{[Color online] Top row: Gaussian white noise current profile. Middle row: Time evolution of membrane potential $V(t)$ with multiplicative noise current and random refractory period $t_{\text{ref}} = 8 + 2\tilde{\eta}(t)$ (ms). Bottom row: Time evolution of membrane potential $V(t)$ with direct input current and direct refractory period $t_{\text{ref}} = 8 $ (ms). The dash line represents the spike threshold $V_{\text{th}} = -55$ (mV).
			%		TThe circuit diagram (left) represents the schematic layout of the integrate-and-firecomponents. The graph (right) depicts voltage vs time for a neuronstimulated by a constant current.
		}\label{fig:1}	
	\end{figure}
\begin{figure}[h!]
		\centering
		\begin{tabular}{ll}
		\includegraphics[width=0.5\textwidth]{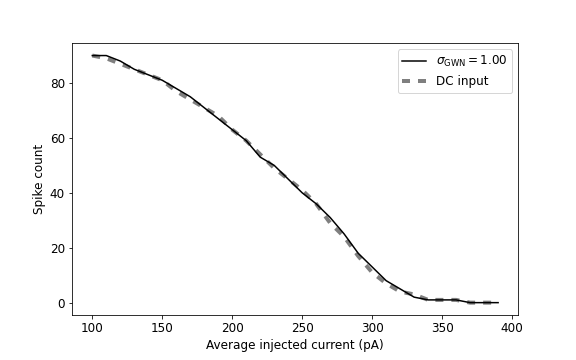} &
		\includegraphics[width=0.5\textwidth]{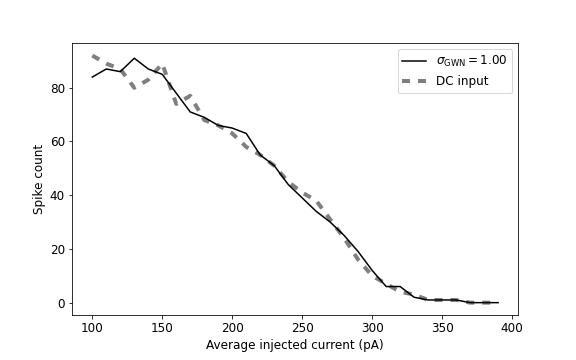} 
			\end{tabular}
		\caption{[Color online] The input-output transfer function of the neuron, output firing rate as a function of input mean for the case with multiplicative noise input current ($\sigma_{\text{Inj}} = 1$).  Left: direct time refractory period $t_{\text{ref}} = 8$ (ms). Right: random refractory period $t_{\text{ref}} = 8 + 4\tilde{\eta}(t)$ (ms).
			%		TThe circuit diagram (left) represents the schematic layout of the integrate-and-firecomponents. The graph (right) depicts voltage vs time for a neuronstimulated by a constant current.
		}\label{fig:1-2}	
	\end{figure}
\begin{figure}[h!]
		\centering
		\begin{tabular}{l}
			\includegraphics[width=0.8\textwidth]{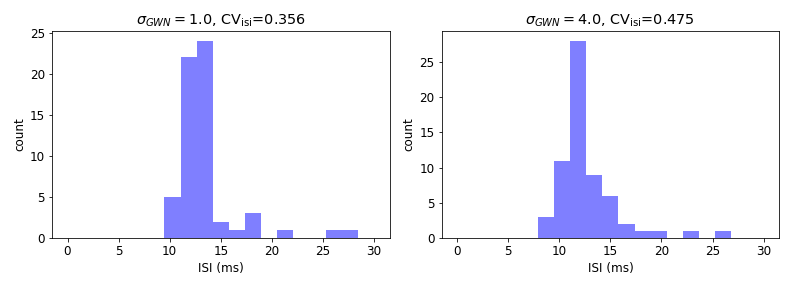} \\
			\includegraphics[width=0.8\textwidth]{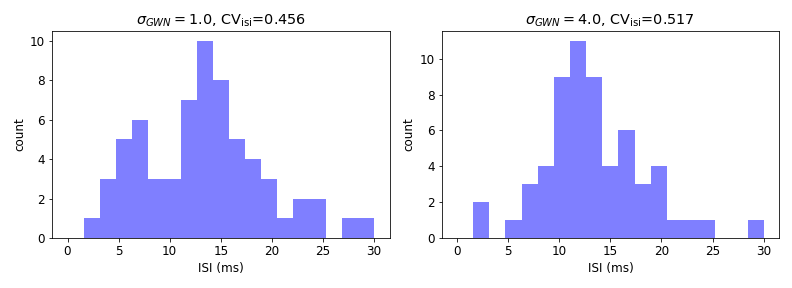}
		\end{tabular}
		\caption{[Color online] ISI histogram distributions for the case with multiplicative noise input current. $t_{\text{ref}} = 8 + 0.5\tilde{\eta}(t)$ (ms). Second row: $t_{\text{ref}} = 8 + 4\tilde{\eta}(t)$ (ms).}
		\label{fig:1-3}
	\end{figure}
\begin{figure}[h!]
		\centering
			\begin{tabular}{ll}
		\includegraphics[width=0.4\textwidth]{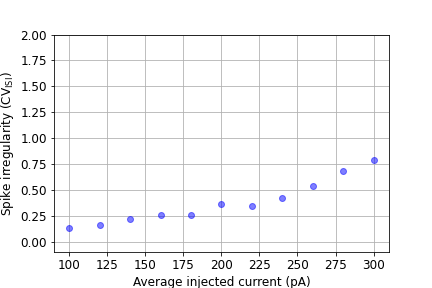} & 	\includegraphics[width=0.4\textwidth]{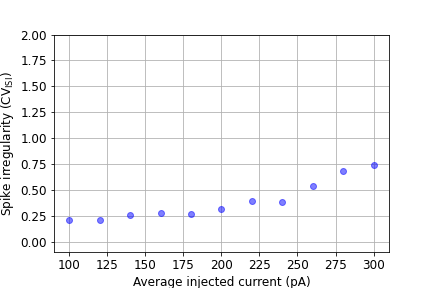}\\
		\includegraphics[width=0.4\textwidth]{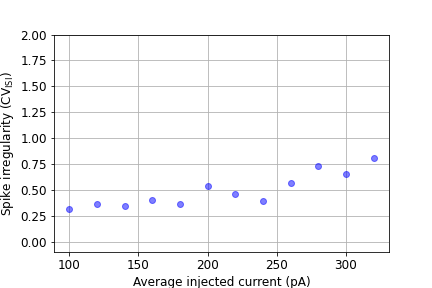} & 	\includegraphics[width=0.4\textwidth]{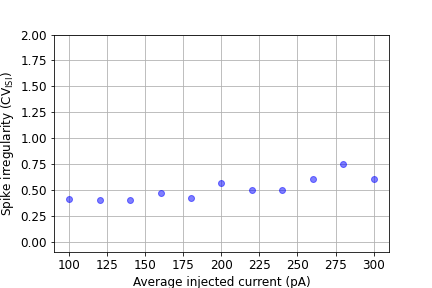}
	\end{tabular}
		\caption{[Color online] Spike irregularity profile in the case with multiplicative noise input current. Top left panel: $t_{\text{ref}} = 8 + \tilde{\eta}(t)$ (ms). Top right panel: $t_{\text{ref}} = 8 + 2\tilde{\eta}(t)$ (ms). Bottom left panel: $t_{\text{ref}} = 8 + 4\tilde{\eta}(t)$ (ms). Bottom right panel: $t_{\text{ref}} = 8 + 6\tilde{\eta}(t)$ (ms).
			%		TThe circuit diagram (left) represents the schematic layout of the integrate-and-firecomponents. The graph (right) depicts voltage vs time for a neuronstimulated by a constant current.
		}\label{fig:1-4}	
	\end{figure}

In Fig. \ref{fig:1}, we have plotted the Gaussian white noise current profile, the time evolution of the membrane potential $V(t)$ with multiplicative noise input current ($I_{\text{inj}} = V(t)(200 + \eta(t))$ (pA)) and direct input current ($I_{\text{inj}} = I_{\text{dc}} = 200V(t)$ (pA)). In this case, we fix the value of $t_{\text{ref}} = 8 + \eta(t)$ (ms). There are bursting moods in the membrane potential in both two cases. This is due to the presence of $V(t)$ in the input current together with the random refractory period in the system. 
 \begin{figure}[h!]
	\centering
	\includegraphics[width=0.7\textwidth]{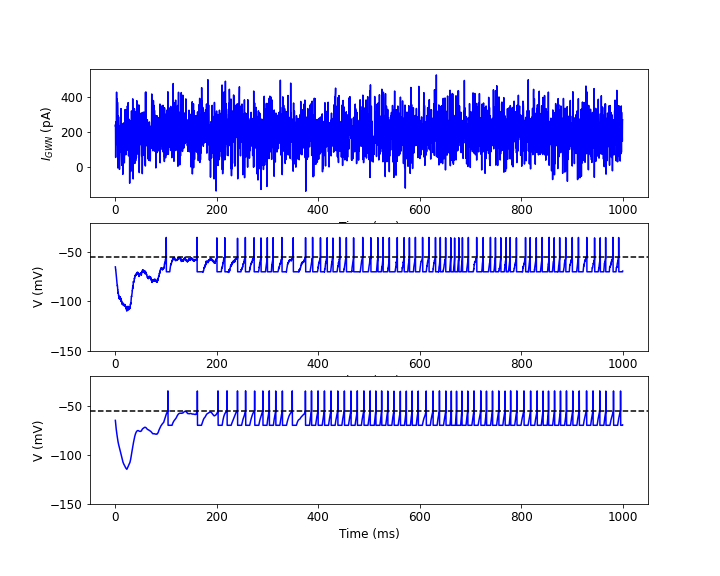} 
	\caption{[Color online] Top row: Gaussian white noise current profile. Middle row: Time evolution of membrane potential $V(t)$ with multiplicative noise current and random refractory period $t_{\text{ref}} = 8 + 2\tilde{\eta}(t)$ (ms). Bottom row: Time evolution of membrane potential $V(t)$ with direct input current and direct refractory period $t_{\text{ref}} = 8 $ (ms). The dash line represents the spike threshold $V_{\text{th}} = -55$ (mV).
		%		TThe circuit diagram (left) represents the schematic layout of the integrate-and-firecomponents. The graph (right) depicts voltage vs time for a neuronstimulated by a constant current.
	}\label{fig:1-1}	
\end{figure}
However, when we increase the leak conductance from $g_L=20$ (nS) to $g_L=200$ (nS), the burst discharges are dramatically reduced in the case with multiplicative noise in Fig. \ref{fig:1-1}. In particular, we observe fluctuations in the membrane potential in the second row of Fig. \ref{fig:1-1}. There is an increase in the time interval between two nearest neighbor spikes in both cases. In order to understand better such phenomena, let us look at the following plots. From now on, we will use the parameter $g_L = 200$ (nS) for cases in Figs. \ref{fig:1-2}-\ref{fig:1-4}. 

In Fig. \ref{fig:1-2}, we look at the input-output transfer function of the neuron, output firing rate as a function of input means. We observe that the input-output transfer function looks quite similar in both cases: direct and random refractory periods. There are slight fluctuations in the spike count profile in the case with random refractory period. It is clear that the presence of multiplicative noise strongly affects the spiking activity in our system compared to the case with additive noise. The spiking activity of the neuron dramatically reduces in presence of the multiplicative noise in the system. 

%If we use a DC input, the F-I curve is deterministic, and we can
%found its shape by solving the membrane equation of the neuron. If we have GWN,
%as we increase the sigma, the F-I curve has a more linear shape, and the neuron
%reaches its threshold using less average injected current.

%1) If we have bigger current fluctuations (increased sigma), the minimum input needed
%to make a neuron spike is smaller as the fluctuations can help push the voltage above
%threshold.
%2) The standard deviation (or size of current fluctuations) dictates the level of
%irregularity of the spikes; the higher the sigma the more irregular the observed
%spikes.	
In Fig. \ref{fig:1-3}, we have plotted the spike distribution as a function of ISI. In the first row of Fig. \ref{fig:1-3}, we see that the spikes are distributed almost entirely in the ISI interval [9;18] (ms). Moreover, when we increase the value of $\sigma_{\text{ref}}$ from 0.5 to 4 the spike irregularity values increase. In addition, they are distributed almost entirely in the ISI interval [1;25] (ms) in the second row of Fig. \ref{fig:1-3}. To understand better such phenomenon we look at the spike irregularity profile of our system in presence of multiplicative noise and random refractory period in Fig. \ref{fig:1-4}. In particular, we observe that the spike irregularity $CV_{\text{ISI}}$ increases when we increase the values of $\sigma_{\text{ref}}$ similar to the case of additive noise. Furthermore, we see that the larger injected currents are, the higher are the values of $CV_{\text{ISI}}$.

Additionally, we notice that the presence of the random refractory period increases the spiking activity of the neuron. The presence of additive and multiplicative noise causes burst discharges in the system. However, when we increase the value of leak conductance, the burst discharges are strongly reduced in the case with multiplicative noise. Under suitable values of average injected current as well as the values of random
input current and random refractory period, the irregularity of spike trains increases. The presence of additive noise could lead to the occurrence of bursts, while the presence of multiplicative noise with random refractory period could reduce the burst discharges in some cases. This effect may lead to an improvement in the carrying of information about stimulating activities in the neuron \cite{Bauermann2019}. Moreover, the study of random factors in the LIF conductance model would potentially contribute to further progress in addressing the challenge of how the active membrane currents generating bursts of action potentials affect neural coding and computation \cite{Kepecs2003}.
	Finally, we remark that noise may come from different sources, e.g., devices, environment, chemical reactions. Moreover, as such, noise is not always a problem for neurons, it can also bring benefits to nervous systems \cite{Faisal2008,Bauermann2019}. 

	\section{Conclusion}
	
	We have proposed and described a LIF synaptic conductance model with random inputs. Using the description based on Langevin stochastic dynamics in a numerical setting, we analyzed the effects of noise in a cell membrane potential. Specifically, we provided details of the models along with representative numerical examples and discussed the effects of random inputs on the time evolution of the cell membrane potentials, the corresponding spiking activities of neurons and the firing rates. Our numerical results have shown that the random inputs strongly affect the spiking activities of neurons in the LIF synaptic conductance model. Furthermore, we observed that the presence of multiplicative noise causes burst discharges in the LIF synaptic conductance dynamics. However, when increasing the value of the leak conductance, the bursting moods are reduced. When the values of average injected current are large enough together with an increased standard deviation of the refractory period, the irregularity of spike trains increases. With more irregular spike trains, we can potentially expect a decrease in bursts in the LIF synaptic conductance system. Random inputs in LIF neurons could reduce the response of the neuron to each stimulus in SNNs and/or ANNs systems. A better understanding of uncertainty factors in neural network systems could contribute to further developments of SNN algorithms for higher-level brain-inspired functionality studies and other applications.
	
	%	Managing such uncertainty factors in neural network systems would allow for improvements of SNNs and/or ANNs systems in the fields as biomedicine and other applications.
	%%%%%%%
	\section*{Acknowledgment}
	Authors are grateful to the NSERC and the CRC Program for their
	support. RM is also acknowledging support of the BERC 2022-2025 program and Spanish Ministry of Science, Innovation and Universities through the Agencia Estatal de Investigacion (AEI) BCAM Severo Ochoa excellence accreditation SEV-2017-0718 and the Basque Government fund AI in BCAM EXP. 2019/00432.

	\bibliographystyle{splncs04}
	\bibliography{mybibn}

\begin{thebibliography}{10}
\providecommand{\url}[1]{\texttt{#1}}
\providecommand{\urlprefix}{URL }

\bibitem{Bauermann2019}
Bauermann, J., Lindner, B.: Multiplicative noise is beneficial for the
  transmission of sensory signals in simple neuron models. BioSystems  178,
  25--31 (2019)

\bibitem{Brigner2020}
Brigner, W.H., Hu, X., Hassan, N., Jiang-Wei, L., Bennett, C.H.,
  Garcia-Sanchez, F., Akinola, O., Pasquale, M., Marinella, M.J., Incorvia,
  J.A.C., Friedman, J.S.: Three artificial spintronic leaky integrate-and-fire
  neurons. SPIN  10(2),  2040003 (2020)

\bibitem{Burkitt2006}
Burkitt, A.: A review of the integrate-and-fire neuron model: I. homogeneous
  synaptic input. Biological Cybernetics  95(2),  97--112 (2006)

\bibitem{Burkitt2006-2}
Burkitt, A.: A review of the integrate-and-fire neuron model: {II}.
  inhomogeneous synaptic input and network properties. Biological Cybernetics
  95(1),  1--19 (2006)

\bibitem{Cavallari2014}
Cavallari, S., Panzeri, S., Mazzoni, A.: Comparison of the dynamics of neural
  interactions between current-based and conductance-based integrate-and-fire
  recurrent networks. Front. Neural Circuits  8(12) (2014)

\bibitem{Chen2021}
Chen, X., Yajima, T., Inoue, I.H., Iizuka, T.: An ultra-compact leaky
  integrate-and-fire neuron with long and tunable time constant utilizing
  pseudo resistors for spiking neural networks. Accepted for publication in
  Japanese Journal of Applied Physics  (2021)

\bibitem{Chowdhury2021}
Chowdhury, S.S., Lee, C., Roy, K.: Towards understanding the effect of leak in
  spiking neural networks. Neurocomputing  464,  {83--94} (2021)

\bibitem{Christodoulou2001}
Christodoulou, C., Bugmann, G.: Coefficient of variation vs. mean interspike
  interval curves: What do they tell us about the brain? Neurocomputing  38-40,
   1141--1149 (2001)

\bibitem{Dayan2005}
Dayan, P., Abbott, L.F.: Theoretical Neuroscience. The MIT Press Cambridge,
  Massachusetts London, England (2005)

\bibitem{Dutta2017}
Dutta, S., Kumar, V., Shukla, A., Mohapatra, N.R., Ganguly, U.: Leaky integrate
  and fire neuron by charge-discharge dynamics in floating-body mosfet.
  Scientific Reports  7(8257) (2017)

\bibitem{Faisal2008}
Faisal, A.D., Selen, L.P.J., Wolpert, D.M.: Noise in the nervous system. Nature
  Reviews Neuroscience  9,  292--303 (2008)

\bibitem{Fardet2020}
Fardet, T., Levina, A.: Simple models including energy and spike constraints
  reproduce complex activity patterns and metabolic disruptions. PLoS Comput
  Biol  16(12),  e1008503 (2020)

\bibitem{Gallinaro2021}
Gallinaro, J.V., Clopath, C.: Memories in a network with excitatory and
  inhibitory plasticity are encoded in the spiking irregularity. PLoS Comput
  Biol  17(11),  e1009593 (2021)

\bibitem{Gerstner2014}
Gerstner, W., Kistler, W.M., Naud, R., Paninski, L.: Neuronal Dynamics: From
  single neurons to networks and models of cognition. Cambridge University
  Press (2014)

\bibitem{Gerum2021}
Gerum, R.C., Schilling, A.: Integration of leaky-integrate-and-fire neurons in
  standard machine learning architectures to generate hybrid networks: A
  surrogate gradient approach. Neural Computation  33,  2827--2852 (2021)

\bibitem{Guo2021}
Guo, T., Pan, K., Sun, B., Wei, L., Y., Zhou, Y.N., W, Y.A.: Adjustable
  leaky-integrate-and-fire neurons based on memristor coupled capacitors.
  Materials Today Advances  12(100192) (2021)

\bibitem{Hendrycks2019}
Hendrycks, D., Dietterich, T.: Benchmarking neural network robustness to common
  corruptions and perturbations. in: International Conference on Learning
  (2019)

\bibitem{Jaras2021}
Jaras, I., Harada, T., Orchard, M.E., Maldonado, P.E., Vergara, R.C.: Extending
  the integrate-and-fire model to account for metabolic dependencies. Eur J
  Neurosci.  54(3),  5249--5260 (2021)

\bibitem{Kepecs2003}
Kepecs, A., Lisman, J.: Information encoding and computation with spikes and
  bursts. Network: Computation in Neural Systems  14,  103--118 (2003)

\bibitem{Latimer2019}
Latimer, K.W., Rieke, F., Pillow, J.W.: Inferring synaptic inputs from spikes
  with a conductance-based neural encoding model. eLife  8(e47012) (2019)

\bibitem{Li2019}
Li, S., Liu, N., Yao, L., Zhang, X., Zhou, D., Cai, D.: Determination of
  effective synaptic conductances using somatic voltage clamp. PLoS Comput Biol
   15(3),  e1006871 (2019)

\bibitem{Mahdi2013}
Mahdi, A., Sturdy, J., Ottesen, J.T., Olufsen, M.S.: Modeling the afferent
  dynamics of the baroreflex control system. PLoS Comput Biol  9(12),  e1003384
  (2013)

\bibitem{Maimon2009}
Maimon, G., Assad, J.A.: Beyond {P}oisson: Increased spike-time regularity
  across primate parietal cortex. Neuron  62(3),  426--440 (2009)

\bibitem{Nandakumar2020}
Nandakumar, S.R., Boybat, I., Gallo, M.L., Eleftheriou, E., Sebastian, A.,
  Rajendran, B.: Experimental demonstration of supervised learning in spiking
  neural networks with phase change memory synapses. Scientific Reports
  10(8080) (2020)

\bibitem{Roberts2017}
Roberts, J.A., Friston, K.J., Breakspear, M.: Clinical applications of
  stochastic dynamic models of the brain, part i: A primer. Biological
  Psychiatry: Cognitive Neuroscience and Neuroimaging  2,  216--224 (2017)

\bibitem{So2012}
So, R.Q., Kent, A.R., Grill, W.M.: Relative contributions of local cell and
  passing fiber activation and silencing to changes in thalamic fidelity during
  deep brain stimulation and lesioning: a computational modeling study. J
  Comput Neurosci  32,  499--519 (2012)

\bibitem{Stiefel2013}
Stiefel, K.M., Englitz, B., Sejnowski, T.J.: Origin of intrinsic irregular
  firing in cortical interneurons. PNAS  110(19),  7886--7891 (2013)

\bibitem{Teeter2018}
Teeter, C., Iyer, R., Menon, V., Gouwens, N., Feng, D., Berg, J., Szafer, A.,
  Cain, N., Zeng, H., Hawrylycz, M., Koch, C., Mihalas, S.: Generalized leaky
  integrate-and-fire models classify multiple neuron types. Nature
  Communications  9(709) (2018)

\bibitem{Kuhn2004}
Teka, W., Marinov, T.M., Santamaria, F.: Neuronal integration of synaptic input
  in the fluctuation-driven regime. The Journal of Neuroscience  24(10),
  2345--2356 (2004)

\bibitem{Teka2014}
Teka, W., Marinov, T.M., Santamaria, F.: Neuronal spike timing adaptation
  described with a fractional leaky integrate-and-fire model. PLoS Comput Biol
  10(3) (2014)

\bibitem{Pottelbergh2021}
{Van Pottelbergh, T. , Drion, G., Sepulchre, R.}: From biophysical to
  integrate-and-fire modeling. Neural Computation  33(3),  563--589 (2021)

\bibitem{Woo2021}
Woo, J., Kim, S.H., Han, K., M.: Characterization of dynamics and information
  processing of integrate-and-fire neuron models. J. Phys. A: Math. Theor.
  54(445601) (2021)

\end{thebibliography}
\end{document}